\documentclass{emulateapj}
\usepackage{epsfig}

\def\simlt{\mathrel{\hbox{\rlap{\hbox{\lower4pt\hbox{$\sim$}}}\hbox{$<$}}}}
\def\simgt{\mathrel{\hbox{\rlap{\hbox{\lower4pt\hbox{$\sim$}}}\hbox{$>$}}}}

\def\vlos{v_{los}}

\shorttitle{dSph Proper Motions} 
\shortauthors{Kaplinghat and Strigari}
\begin{document}

\title{Proper Motion of Milky Way Dwarf Spheroidals from Line-of-Sight Velocities}

\author{Manoj Kaplinghat and Louis E. Strigari}
\email{mkapling@uci.edu,lstrigar@uci.edu}
\affil{Center for Cosmology,
Department of Physics \& Astronomy, University of California, Irvine,
CA 92697} 



\begin{abstract}
Proper motions for several Milky Way dwarf spheroidal (dSph) galaxies 
have been determined using both ground and space-based imaging. These
measurements require long baselines and repeat observations and
typical errors are of order ten milli-arcseconds per century.   
In this paper, we utilize the effect of ``perspective rotation" to
show that systematic proper motion of some dSphs can be determined to
a similar precision using {\em only} stellar line-of-sight velocities.
We show that including the effects of small
intrinsic rotation in dSphs increases the proper motion errors by about a 
factor of two. 
We provide error projections for future data sets,
and show that proposed thirty meter class telescopes will measure the
proper motion of a few dSphs with milli-arcsecond per century
precision. 
\end{abstract}


\keywords{Cosmology: dark matter, theory--galaxies: kinematics and dynamics--Astrometry}

 \maketitle

\section{Introduction} 
The three-dimensional motions
of stars and galaxies provide valuable information on the local
Universe, ranging
from planetary companions of nearby stars to the orbital
properties of nearby galaxies and Galactic satellites
~\citep{Unwin:2007wj}. 
However, because measurements of proper motion 
must be done with respect to cosmic standards of rest, 
such as background galaxies and quasi-stellar objects, 
they depend sensitively on the number and nature of background 
objects in the target field. 
To obtain measurements of required precision, 
long baselines and repeat measurements are necessary, 
ranging from several years for space-based
telescopes to tens of years for ground-based telescopes. 

Nonetheless, despite their low luminosities, large distances, and 
small angular separations,  about a dozen nearby galaxies 
now have measured proper motions~\citep{Piatek_Fornax}. 
Among the nearest of these galaxies 
are the dwarf spheroidals,
which have observed luminosities that 
vary anywhere from a 
thousand to a million times the luminosity of the Sun. 
These galaxies are supported primarily by their  
velocity dispersion and have high mass-to-light ratios $\simgt 100$
M$_\odot/$ L$_\odot$ 
~\citep{Mateo1998,Lokas2005,Gilmore2007,StrigariRedefining,
SimonGeha,Walkeretal2007}.
Proper motions have been measured for several Milky 
Way dSphs~\citep{Piatek_UrsaMinor,Piatek_Sculptor,Piatek_Fornax}, 
with errors $\sim 10$ milli-arcseconds per century, corresponding to
transverse velocity errors $\sim 100$ km s$^{-1}$ for typical
dSph distances. In contrast, their velocities in the direction of the
observer are now known to $\simlt 1$ km s$^{-1}$.  

In this paper, we introduce an alternative technique for 
determining the proper motions of Milky Way dSphs.  
Specifically, we utilize present samples of stellar line-of-sight
velocities together with the effect known as ``perspective rotation," 
in which the tangential motion of the galaxy contributes
to the measured line-of-sight velocity at 
large angular separations from the center of the galaxy. 
Perspective rotation has been detected in Galactic globular 
clusters~\citep{Merritt1997}, and 
has been used to measure the distance to the Large Magellanic Cloud
~\citep{Gould2000} and the mass of M31~\citep{vanderMarel:2007yw}. 
Here we show that, using perspective rotation, proper motions of several
dSphs can be determined to a precision rivaling the best 
existing ground and space-based measurements. 
Additionally, the first proper motion measurements will be possible
for several dSphs that are difficult to access via traditional
methods.   

\section{Perspective Rotation} 
Perspective rotation is simple to understand if we consider the dSphs
as extended objects.  Because of their close proximity and spatial
extent,  the line-of-sight velocities of the stars vary as a function
of projected radial separation, $R$, from the center 
of the galaxy. As $R$ increases, the line-of-sight velocities
receive increasingly larger contributions
from the tangential motion of the object
in space. The net result of this radially varying line-of-sight
velocity is known as perspective rotation
~\citep{Feast1961}. 
An object that is not rotating intrinsically acquires
a velocity gradient that is proportional to the 
projected distance from the center of the galaxy. 

To describe the effect of perspective rotation, 
we define a cartesian coordinate system in which the $z$-axis
points in the direction of the observer from the center of the galaxy,
the $x$-axis points in  the direction of decreasing right ascension,
and the $y$-axis  points in the direction of increasing declination.  
The angle $\phi$ is measured counter-clockwise from the positive
$x$-axis, and $\rho$ is the angular separation from the center of the
galaxy.  The line-of-sight velocity is then  
\begin{equation} 
\vlos = v_x \sin \rho \cos \phi + v_y \sin \rho \sin \phi
- v_z \cos \rho. 
\label{eq:vlos}
\end{equation}

For all of the dSphs that we consider, it is appropriate
to use the small angle approximation, 
$\sin \rho \simeq R/D$, where $R = \sqrt{x^2 + y^2}$, 
and $D$ is the
distance to the dSph. Then using $\sin \phi = y/R$,
equation~\ref{eq:vlos} can be  written as $\vlos = v_x x/D + v_y y/D -
v_z$. In the limit that $\sin \rho \ll 1$, the line-of-sight velocity  
is constant across the dSphs and we recover in this limit that $\vlos
= -v_z$. It is evident from equation~\ref{eq:vlos}  that the
transverse velocities $v_x$ and $v_y$ have  the maximal contributions
for galaxies that are the appropriate combination of the most nearby
and the most spatially extended. 

\section{Likelihood Function and Error Projections} 
We define the likelihood function for an observed set of
line-of-sight velocities as
\begin{equation} 
{\cal L}(\vec{\theta}) = \prod_{i=1}^N
\frac{1}
{\sqrt{2\pi\sigma_i^2}}
\exp\left[-\frac{1}{2}
\frac{(v_i-{\vlos}_{,i})^2}{\sigma_i^2}\right]. 
\label{eq:likefull}
\end{equation} 
Here $\vec{\theta}$ is the set of parameters that describe 
the model of the galaxy. 
The product is over the total number of stars, $N$, with measured
velocities, and $i$ is an index that represents a star that is 
located at a fixed projected position. 
The total velocity dispersion, $\sigma$, is the sum of the intrinsic
dispersion, $\sigma_{los}$, and the dispersion from the measurement,
$\sigma_m$.  

We determine $\sigma_{los}$ from standard dynamical equilibrium
analysis, assuming that the potentials  
of the dSphs are spherically-symmetric.
The Jeans equation for the radial velocity dispersion $\sigma_r$ is 
\begin{equation}
r \frac{d(\rho_{\star} \sigma_r^2)}{dr} =  - \rho_{\star}(r) 
\frac{GM(r)}{r}- 2 \beta(r) \rho_{\star}(r) \sigma_r^2.
\label{eq:jeans}
\end{equation}
Here $M(r)$ is the halo mass, 
$\beta(r) = 1 - \sigma_\theta^2/\sigma_r^2$ 
is the stellar velocity anisotropy, and $\rho_\star(r)$ is the three-
dimensional density for the stars, which is determined from the 
measured surface density of stars, $I(\rho)$. For $I(\rho)$, we
use King profiles, which are good fits to the surface densities
of the dSphs that we examine~\citep{King1962,Irwin1995}. 
King profiles are fit by a ``core" radius, $r_c$, and a limiting radius, 
$r_{\lim}$. For a King profile the limiting radius is the tidal 
radius of the stars.  

The radial velocity dispersion can be determined by solving
for $\sigma_r$ using equation~\ref{eq:jeans} and imposing 
the boundary condition $\sigma_r \rightarrow 0$ as
$r \rightarrow \infty$. We can then use 
equation~\ref{eq:jeans} and the definition of $\beta$
to convert the velocity dispersions implied from 
equation~\ref{eq:vlos} into observable
quantities by integrating along the line-of-sight through the dSph. 
Performing the integration gives
\begin{equation}
\sigma_{los}^2(\rho) = \frac{2}{I(\rho)} 
\int_{r_{min}}^{\infty} 
\left[1 - \beta (r) \alpha^2 \right]
\frac{\rho_{\star}(r) \sigma_{r}^{2} r dr}{\sqrt{r^2-D^2\sin^2 \rho}}. 
\label{eq:1projected} 
\end{equation}
Here, $r_{min} = D\tan \rho$; $\alpha^2 \equiv  \cos^2 [\theta(r)] \sin^2 \rho
+ 2 \sin [\theta(r)] \sin \rho \cos [\theta(r)] \cos \rho 
+ \sin^2 [\theta(r)] \cos^2 \rho$, with  
$\cos [\theta(r)] = (D-D\cos^2\rho + \sqrt{r^2-D^2\sin^2\rho}\cos \rho)/r$. 
For simplicity we assume a constant $\beta$; we find that 
our results below do not depend on whether $\beta$
is a fixed constant or is allowed to be a more complicated function of radius. 
Accounting for the definition of the line-of-sight velocity
in equation~\ref{eq:vlos}, equation~\ref{eq:1projected} 
differs from the standard definition of the projected 
velocity dispersion at a fixed $R$~\citep{binney87}.  

We take the dSphs to be dark matter dominated, 
with halos described by the density profile 
$\rho(r)= \rho_0(r/r_0)^{-a}(1+r^b/r_0^b)^{(a-c)/b}$. 
We take the slopes, $a$, $b$, and $c$, the
scale density $\rho_0$, and $r_0$ to be 
unknown parameters and marginalize over 
them with uniform priors. Our results are 
independent of the halo mass model, provided
the projected velocity dispersion is fixed
to match the observed, nearly  
flat dispersion profiles of the dSphs~\citep{Walkeretal2007}. 

We are interested in using the likelihood function 
in equation~\ref{eq:likefull} in concert with 
equations~\ref{eq:jeans} and~\ref{eq:1projected}
to project the errors on the components of the 
transverse velocity, $v_x$ and $v_y$. The attainable errors depend  on 
the covariance matrix for the model parameters
$\vec{\theta}$, which we will approximate 
by the Fisher information matrix  
$F_{ab} = \langle \partial^2 \ln {\cal L} /
\partial \theta_a \partial \theta_b  \rangle $ \citep{kendallstuart69}. 
Our model parameters are those 
that describe the dark matter halo, 
the velocity anisotropy, and  
the spatial motion of the galaxy. 
The inverse of the Fisher matrix, ${\bf F}^{-1}$, 
provides an estimate of the covariance between the parameters, 
and $\sqrt{F_{aa}^{-1}}$ approximates the error 
on the parameter $\theta_a$. The Cramer-Rao inequality
guarantees  that $\sqrt{F_{aa}^{-1}}$ is the minimum
possible variance on the $a^{th}$  parameter for an unbiased
estimator. 
Using ${\bf F}^{-1}$ in place of the true covariance matrix involves
approximating the likelihood function of the parameters as Gaussian
near its peak, so  ${\bf F}^{-1}$ will be a good approximation
to the errors on parameters that are well-constrained. 

We construct the Fisher matrix by differentiating the log 
of the likelihood function in equation~\ref{eq:likefull}, 
and averaging over the data. 
Performing the appropriate averaging, and using the above
definition of the line-of-sight velocity in equation~\ref{eq:vlos},
our final expression for the Fisher matrix is 
\begin{equation} 
F_{ab} 
= \sum_{\imath=1}^{N} \left(
\frac{1}{\sigma_\imath^2}\frac{\partial {\vlos}_{,\imath}}{\partial \theta_a}
\frac{\partial {\vlos}_{,\imath}}{\partial \theta_b}
+\frac{1}{2}
\frac{1}{\sigma_\imath^4}\frac{\partial \sigma_{los,\imath}^2}{\partial \theta_a}
\frac{\partial \sigma_{los,\imath}^2}{\partial \theta_b}\right).  
\label{eq:fisher} 
\end{equation} 
In deriving equation~\ref{eq:fisher}, we have assumed no 
correlations between the theory and the measurement 
dispersions. 

In the second term in equation~\ref{eq:fisher}, the derivatives 
are with respect to the theory dispersion alone, whereas both 
of the contributions to the variance sum in the denominator. 
For the classical, well-studied dSphs, the intrinsic velocity 
dispersions are of order  5-15 km s$^{-1}$, while the 
mean measurement uncertainty is less than 2 km s$^{-1}$, 
so the dominant contribution to the variance comes
from the theoretical distribution function (For many of 
the newly-discovered satellites, however, both contributions to 
the dispersion are similar~\citep{SimonGeha}). 
Here we are interested in the highest luminosity dSphs, 
so in all of our examples the error from the 
intrinsic dispersion of the dSph is the dominant contribution. 

Equation~\ref{eq:fisher} shows that, to determine the error 
on any of the $\vec{\theta}$ parameters, we need two 
pieces of information: 1) the distribution of stars within the 
dSph that have measured velocities, and 2) the error on the 
velocity of each star. The projected errors are independent 
of the mean velocity of the stars. The errors on the parameters
describing the dark matter halo and the velocity anisotropy 
enter only through the second term in equation~\ref{eq:fisher} 
when we differentiate the velocity dispersions in 
equation~\ref{eq:1projected}. The errors on the three 
velocity components $v_x$, $v_y$, and $v_z$ are independent
of the parameters that describe the halo and velocity 
anisotropy. 

\section{Results} 
To project errors on the proper motion components, 
we draw stars from uniform distributions for both the
$x$ and $y$ spatial coordinates. We assign a measurement
error of $\sigma_m = 2$ km s$^{-1}$ for each star, 
which represents a conservative upper limit for 
the mean error in high 
luminosity dSphs. Our results
are not strongly sensitive to the shape of the distribution
from which we draw stars, except in the case that the
stars are strongly centrally-concentrated within the 
King core radius. 

We consider two fiducial models for dSphs. The
first model is a compact, nearby system 
designed to represent Draco, which is located
at a distance $D = 80$ kpc and is described
by a King core and limiting radius of $0.18$
and $0.93$ kpc, respectively. From ground 
based measurements, the proper
motion of Draco is 
$(\mu_\alpha, \mu_\delta)=(60 \pm40,110\pm50)$ 
mas per century
~\citep{ScholzIrwin1994}. We note that Sculptor is 
at a distance similar to Draco, has 
a higher luminosity, and contains a similar 
number of line-of-sight velocities,  
so it may also present an 
interesting target. However, as we discuss below, 
the prospects for measuring its proper motion 
using our methods may be complicated
by the presence of intrinsic
rotation~\citep{Battaglia:2008jz}. 

As our second
model, we consider a more distant and more
extended dSph, designed to represent
Fornax ($D = 138$ kpc). Fornax is described by a King
core radius and limiting radius of $0.40$
and $2.7$ kpc, respectively. The proper
motion of Fornax from {\em Hubble Space Telescope (HST)} 
observations 
is $(\mu_\alpha, \mu_\delta)=(47.6\pm 4.6,-36.0\pm 4.1)$ 
mas century$^{-1}$~\citep{Piatek_Fornax}. 
At present, Fornax has a total of $\sim 2000$
measured line-of-sight velocities
~\citep{Walkeretal2007}, and 
$\sim 8600$ red giant
stars greater than 20th magnitude for which
future observations will be possible. 

Examining equation~\ref{eq:fisher}, the
projected errors on parameters generally 
depend on the fiducial point in  
parameter space. We consider fiducial models
fit to match the
projected velocity dispersions of Fornax and Draco. 
We find that $\rho_0 = 10^7$ M$_\odot$ kpc$^{-3}$, 
$r_0 = 2$ kpc, $a = 1$, $b=1$, and $c=3$ provides
a good description of their dispersion profiles
~\citep{Strigari:2006rd}. For Fornax
we take the stellar distribution to be isotropic, $\beta = 0$, 
while for Draco we take a tangential anisotropy of $\beta = -1$. 
The dark matter halos have a mass of $\sim 10^7$ M$_\odot$
within their inner 300 pc.

\begin{figure}
\plotone{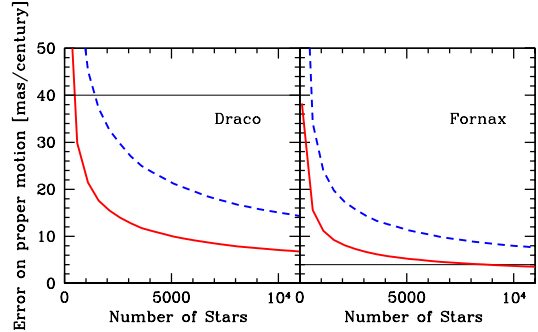}
\caption{The projected errors on the proper motion of dSphs as
a function of the number of line-of-sight velocities, 
for two different models. The {\em left} panel uses the 
Draco fiducial model, while the {\em right} panel 
uses the Fornax fiducial model. In each panel, 
the solid (red) curve is for a non-rotating system, 
and the dashed (blue) curve includes a small rotational
component. The lines indicate the approximate present errors on the
proper motions. 
Measurement errors of 2 km s$^{-1}$ are assumed
for line-of-sight velocities. 
\label{fig:perspective_rotation}
}
\end{figure}

In Figure~\ref{fig:perspective_rotation} we show 
the projected errors
on the proper motion components as a function 
of the number of line-of-sight velocities. For each model, 
for numbers of stars $\simgt 1000$,  
the errors reduce to tens of milli-arcseconds per century, 
rivaling the best measurements from {\em HST} 
~\citep{Piatek_UrsaMinor,Piatek_Sculptor,Piatek_Fornax}.
For the specific case of Draco, we project that the current 
samples of line-of-sight velocities~\citep{Walkeretal2007}
will reduce the error on its proper motion by a factor
of at least two relative to the best ground-based 
measurements.    
For each curve in Figure~\ref{fig:perspective_rotation}, we
show only one component of the proper motion, 
in this case the component corresponding to the transverse velocity $v_x$, 
and note that the constraints
on the component corresponding to $v_y$ are similar by symmetry.  
We also find $v_z$ to be strongly constrained, $\simlt 1$ 
km s$^{-1}$ for the samples of stars that we consider. 

It is important to note that the projected errors in 
Figure~\ref{fig:perspective_rotation} are valid only for
the spatial distributions of stars and the measurement
errors that we consider.  
The true constraints will depend on the detailed distributions
of these quantities, and may in fact be reduced with more 
precise velocity measurements or varying stellar spatial 
distributions. For example, using the sample of Fornax
red giants greater than 20th magnitude, we find 
sensitivities reduce to $\sim 3$ mas per century. 

\section{Contribution from Intrinsic Rotation} 
In the analysis above, we have assumed a negligible
contribution from the intrinsic rotation in dSphs. 
This is a good model for the majority of these
systems, which exhibit no detectable 
rotational or streaming motions in the kinematic
data~\citep{Walker2006,Koch2007} 
(See however the recent results of 
~\cite{Battaglia:2008jz}, which show that
rotation may be present in Sculptor). 
 
Though the intrinsic rotation in dSphs is small, 
it is important to determine how even a small
signal may degrade the constraints
on the proper motions we have determined above. 
Probably the simplest rotational model to consider is a
sinusoidal variation in the rotational velocity
as a function of azimuthal angle~\citep{Drukier1998}. 
In this model, a term of the form $A\sin(\phi_i + \phi_0)$
can then be added to equation~\ref{eq:vlos}, where $A$ 
is the amplitude of the rotational motion, and
$\phi_0$ is the projected axis about which the rotation
occurs. 
We note that higher order multipoles may exist in the velocity field
as a result of more complicated rotation or tidal effects; however
here we assume  these  higher order terms are sub-dominant to the
leading dipole term.  

We add the parameters $A$ and $\phi_0$ to the 
Fisher matrix, and marginalize over them with uniform 
priors. In Figure~\ref{fig:perspective_rotation}, we show 
the projected proper motion errors  including the effect of 
intrinsic rotation. Accounting for intrinsic rotation, we find 
that the errors on the proper motions may increase by 
about a factor of two for a fixed number of stars. Again, the 
details of the constraints depend on the exact distribution 
of the stars. Regarding the rotational parameters themselves, 
we find that $\phi_0$ is not well-constrained, but $A$ is determined 
to a precision of $\simlt 1$ km s$^{-1}$ with $\simgt 500$ line-of-sight 
velocities. Similar to $v_x$ and $v_y$, these errors on $A$ are 
independent of its mean value. We note that for Draco and Fornax 
the data suggest $A < 1$ km s$^{-1}$. 

We reiterate that our parametrization of rotation 
 is a toy model introduced to understand the effect of including
 rotation on the precision with which proper motion may be measured. 
Although we have
determined the errors on $A$ by solving for $\sigma_{los}$
using the Jeans equation, this procedure is not self-consistent,
 because in the presence of rotation the Jeans equation itself must be 
 modified. In a rotating system,  this would be important for
 determining the  parameters describing the halo or the velocity
 anisotropy.   However, since here we are interested only in  
determining the velocity components, all that 
we demand is $\sigma_{los}$ have enough freedom at each 
projected position to fit the kinematic data. 
As long as the streaming motion is small, as 
in the case of the dSphs we consider, this procedure 
should provide a good approximation to the 
errors on the model parameters $v_x$, $v_y$, $v_z$, and $A$. 

\section{Conclusions} 
We have shown that line-of-sight velocities from dwarf
spheroidals can determine the transverse velocities of these galaxies
to a precision of $\simlt 100$ km s$^{-1}$. This measurement 
utilizes ``perspective rotation," or the variation of the line-of-sight 
velocity across the galaxy resulting from its proper motion. 
The above sensitivity is similar to the current 
measurements from {\em HST} and ground-based imaging
for dSphs such as Fornax, Carina, and Sculptor.  
Using perspective rotation, 
the proper motion errors for several dSphs, including Draco, 
will be reduced by at least a factor of two, and proper motions for dSphs
such as Sextans, Leo I, and Leo II will be determined for the
first time. We find that, with $\sim 10^3$ stars, the proper motion of 
Sextans can be determined to a sensitivity of $\sim 5$
mas per century, while for Leo I and Leo II we project 
errors of $\sim 10 $ mas per century with $10^3$ stars. 

Prospects are promising for improving on these 
measurements with future, larger samples of line-of-sight velocities. 
For example, we project that with $\sim 10^4$ line-of-sight
velocities from Fornax, the errors are reduced by 
$\sim$ 40\%. Proposed
thirty meter class telescopes, such as
the Thirty Meter Telescope (TMT) (http://www.tmt.org) 
and the Giant Magellan Telescope (GMT) (http://www.gmto.org), 
may measure velocities for
$\sim 10^4$ stars in multiple dSphs, obtaining
milli-arcsecond per century sensitivity. 
Nearby ultra-faint dwarfs, with extent $\sim 100$ pc 
and distances $\simlt 50$ kpc (Coma Berenices, Willman 1, 
Ursa Major II), may also be studied. In these objects
we project errors $\sim 100$ mas per century with 
the optimistic scenario of $\sim 10^3$ stars.  
 
 A determination of dSph proper motions will provide
 a full three-dimensional mapping of their orbits, 
 with prospects of improving the measurement 
of the Milky Way dark matter halo mass~\citep{LittleTremaine}. 
 In addition, the proper
 motions will allow for a detailed comparison 
 to the orbital properties of dark matter subhalos in numerical
 simulations~\citep{Diemand:2006ik}, and provide 
 information on the merger and accretion histories
 of these satellites. It will be possible to determine
 which of the satellites are bound to the Milky Way, 
a property that may be closely intertwined with the 
complex star formation histories of several satellites 
~\citep{Besla:2007kf,Mateo:2007xh}.  
 
 \section{Acknowledgments} 
 We are grateful to Matt Walker for several important discussions, and 
 for providing us with positions of Fornax 
 red giants. We thank Betsy Barton, James Bullock, Jeff Cooke, 
 Marla Geha, Josh Simon, and Beth Willman for discussions and encouragement.
 We acknowledge support from NSF grant
AST-0607746.  

\bibliography{perspective}
    
\end{document}